\documentstyle[preprint,aps,epsfig]{revtex}
\begin{document}
\title{On the tritium $\beta $- electrons energy shape}
\author{ V.I. Savichev}
\address{Fakult\"at f\"ur Physik, Universit\"at Freiburg, 
Hermann-Herder-Strasse 3, D-79104 Freiburg, Germany }
\date{\today}
\maketitle
\begin{abstract}
A series of high presicion atomic experiments was carried out last decade to get 
a better  estimate for the electron (anti) neutrino mass $m_\nu$. The reaction 
to be observed is molecular tritium $\beta$- decay.
The $m_\nu$ value serves as one of the tune  parameters to fit $\beta$- electrons spectrum
via reference theoretical one.
The unexpected message consists of  that parameter 
${m^2_\nu}$ has to be fixed at some negative value to get best statistically 
reliable fit. Apart of some exotic scenarios suggested the problem remains open. 
In this paper we tried to reanalyze ground features of  the 
final states spectrum (FSS) and its influence on the $\beta$- electrons spectrum. A new 
approach has been developed which gives some hints to make proper modifications 
to the tabulated FSS in the course of experimental fits.
\end{abstract}


\section{Introduction}

The last atomic neutrino experiments 
base on the observation of molecular tritium $\beta$- decay \cite{exp} :
\begin{equation}
T_2 \rightarrow  T ^3He^+ + e^-  + \tilde{\nu}_e
\label{eq:1}
\end{equation} 
The high energy edge of $\beta$- electrons spectrum corresponds to  about  18.6 KeV.
The sudden change of the nuclear charge and hard recoil kick lead to the population 
of the  hundreds  electronic- rovibrational states of final $T ^3He^+$ molecule.
The neutrinos $\tilde{\nu}_e$ are hardly observable but they influence 
total energy sharing and consequently $\beta$- electrons energy spectrum.

The $\beta$- electrons differential energy spectrum can be written 
in the  following form \cite{th1}:
\begin{equation}
\begin{array}{rcl}
\left| \frac{d N(\epsilon_\beta)}{d \epsilon_\beta}\right| & = & A F(p_\beta,Z) E_\beta p_\beta 
\sum_n P_n(p_\beta) \epsilon_n  \left({\epsilon_n^2- {m^2_\nu} c^4}\right)^{1/2}
\theta (\epsilon_n- {m_\nu} c^2)   ,  \\
 & & \quad \epsilon_n = W_0-\epsilon_\beta -E_n, \epsilon_\beta = E_\beta  - m_e c^2 .
\end{array}
\label{eq:2}
\end{equation} 
The sum in (\ref{eq:2}) runs over final states of  $T ^3He^+$ molecule;
$E_\beta, \epsilon_\beta$ are the total relativistic-  and kinetic energies  and 
$p_\beta$ is the momentum of $\beta$-electrons; $F(p_\beta,Z)$ is the so called Fermi factor 
which takes into account Coulomb charge $Z$ of the daughter nucleus \cite{Konopinski}. 
If we will count final energies 
$E_n$ from the ground $E_g$ state of $T ^3He^+$, the end-point  
$\beta$ -electron kinetic energy $W_0$ is given by the relation:
\begin{equation}
W_0 = (M_t-M_{\alpha})c^2 + E_{T_2} - E_g - E_R - m_e c^2,  
\label{eq:3}
\end{equation}    
where $E_{T_2}$ is the ground state of parent $T_2$ molecule; 
$E_R= \frac {p^2_\beta}{4 M_t}$ is the molecular center of mass recoil energy. 
The final states population probabilities $P_n(p_\beta)$ are given by the 
corresponding matrix elements:
\begin{equation}
P_n(p_\beta) = \int \frac{d \hat{\bbox{q}}}{4\pi} |
 < n | \displaystyle{ e^{i (\bbox{q},\bbox{R})}} | T_2>|^2  
 \quad , \quad \bbox{q} = \frac{\bbox{p}_\beta}{2} ,  
\label{eq:4}
\end{equation}
where $\bbox{R}$ is the internuclear distance \cite{foot1}. 
The  $\beta$- spectrometers currently operating in Mainz and Troitsk 
 detect actually integral spectrum for which 
we have : 
\begin{equation}
\begin{array}{rcl}
N(\epsilon_\beta) & = & \frac{A}{3} F(p_\beta,Z) E_\beta p_\beta 
\sum_n P_n(p_\beta)  \left({\epsilon_n^2- {m^2_\nu} c^4}\right)^{3/2}
\theta (\epsilon_n- {m_\nu} c^2)    .
\end{array}
\label{eq:5}
\end{equation} 
The standard procedure consists of fitting  experimental spectrum close to the 
end-point  energy $W_0$ using formula 
(\ref{eq:5}) varying three parameters  $A, E_0, m^2_\nu$   and value of background.

Since first quantitative results have been published in the literature, immense work 
has been done to purify theoretical spectrum in many different aspects 
\cite{th2}. They showed that about 99\% of final states spectrum covers over 100 eV. It 
mostly consists of the several eV narrow rovibrational multiplets wired on bound and resonance 
adiabatic electronic states as it is shown in Fig.\ \ref{fig1}.  
Some 57\% falls on to electronic ground state of $T ^3He^+$. The  rovibrational multiplet 
consists of  hundreds bound and resonance states distributed "...unevenly in a somewhat 
erratic manner" with average rotational quantum number $J \sim 22\div 25$  \cite{th1}.
The reason is a huge, on the atomic scale, recoil momentum $q \sim 18$ releasing.

The results of extensive joint fits between experiments  and available  
theoretical tables show with a certainty that parameter 
$m^2_\nu$  should be set at a {\it negative} value. The definite  
value explicitly depends  on the chosen energy fit interval. Though physically 
unsatisfactory, it permits to draw up an upper estimate $m_\nu \le 2 \div 3$ 
$eV/c^2$. 

The critical values of $m_\nu \sim 1$ $eV/c^2$ will be 
decisive, e.g. for "dark matter" problem. The series  of experiments aiming to hit this 
border are on the list of Mainz-Troitsk collaboration. The "negative $m^2_\nu$" problem 
will represent  then one of the serious bottlenecks. To exclude reasons related to the FSS
we went to reanalyze basic issues in the theoretical final states spectrum and its influence 
on the $\beta$ -electrons spectrum.

\section{$\beta$-spectrum: general features.}

As Eqs.(\ref{eq:2}), (\ref{eq:5}) suggest, the sign of non-zero neutrino mass is the non-analytic
behavior close to the branch energy of given channel. We would like to show, however, that  
a somewhat simpler functional form is enough for all practical purposes. Let us take for 
granted that    $m^2_\nu$ is a small parameter:
\begin{equation}
\epsilon_n^3- {m^2_\nu} c^4 \epsilon_n + O \left(\left|\frac{({m^2_\nu}c^4)^2}
{\epsilon_n}\right|\right) ,
\label{eq:5a}
\end{equation}
which gives for the  spectral sum itself:
\begin{equation}
\begin{array}{rcl}
\sum_n P_n(p_\beta)  \left[\epsilon_n^3- {m^2_\nu} c^4\epsilon_n \right]
\theta (\epsilon_n- {m_\nu} c^2)    .
\end{array}
\label{eq:5b}
\end{equation}
We can further expand $\theta$-function 
\begin{equation}
\theta (\epsilon_n- {m_\nu} c^2) = \theta (\epsilon_n) -\delta (\epsilon_n){m_\nu} c^2 + 
\ldots 
\label{theta_exp}
\end{equation}
to show that neutrino mass term enters into the spectrum, essentially, via linear 
energy term (\ref{eq:5b}). We plot in Fig. \ref{fig2} absolute difference between 
spectral sum in (\ref{eq:5}) and its linear approximation (\ref{eq:5b}) 
using FSS from  \cite{th1}. It shows that absolute difference  
is uniformly bounded and decreases roughly as $({m_\nu} c^2)^4/(W_0-\epsilon_\beta)$. 
The formula (\ref{eq:5b}) can be 
further transformed to the particularly transparent form:
\begin{equation}
\begin{array}{rcl}
 P_\epsilon \left[\epsilon^3- 3 <E_n>_\epsilon \epsilon^2 + 3  <E^2_n>_\epsilon \epsilon 
 - \frac{3}{2} {m^2_\nu} c^4 (\epsilon -  <E_n>_\epsilon) -  <E^3_n>_\epsilon \right],
\end{array}
\label{eq:5c}
\end{equation}
where we introduced FSS cumulative momenta 
$$ P_\epsilon = \sum_n \theta (\epsilon_n), 
 \quad <E_n>_\epsilon = 1/P_\epsilon\sum_n E_n\theta (\epsilon_n) \ \ldots
$$
It shows that $\beta$-spectrum can be parametrised using few statistical characteristics
of the full FSS.

\section{$\beta$-spectrum: operator formulation. }

To find some  way to calculate cumulative momenta directly, we use in (\ref{eq:2}) well-known 
substitution $E_n \rightarrow H$, where $H$ is Hamiltonian of  $T ^3He^+$ molecule. 
Completing the sum over final spectrum, we get:      
\begin{equation}
\begin{array}{l} 
N(\epsilon_\beta)  =  \frac{A}{3} F(p_\beta,Z) E_\beta p_\beta 
\int \frac{d{\hat{\bf q}}}{4\pi} <T_2| \displaystyle{e^{ - i ({\bf q},{\bf R})}} 
\left( \hat{\epsilon}^2 -  m^2_\nu c^4\right)^{3/2}    
 \theta (W_0-\epsilon_\beta - H)  
  \displaystyle{e^{ i ({\bf q},{\bf R})}} |T_2> ,  
\end{array} 
\label{eq:7}
\end{equation} 
here $\epsilon = W_0-\epsilon_\beta -H$.
It is easy to justify the following commutation relation:
\begin{equation} 
H \displaystyle{e^{ i ({\bf q},{\bf R})}} = \displaystyle{e^{ i ({\bf q},{\bf R})}}
 \left[ H + \frac{q^2}{2 M } + \frac{({\bf q},{\hat{{\bf p}}}_{{\bf R}})}{M} \right] ,
\label{eq:8}
\end{equation}
using explicit form of kinetic energy operator $T_k = -\frac{1}{2M}\Delta$. We  
can eliminate now recoil exponent, making simultaneous substitution in (\ref{eq:7}):
\begin{equation}
W_0-\epsilon_\beta -H \rightarrow    W_0-\epsilon_\beta -\frac{q^2}{2 M } -
  \frac{({\bf q},{\hat{{\bf p}}}_{{\bf R}})}{M} - H
\label{eq:9}
\end{equation} 
The term $\frac{q^2}{2 M }$ can be interpreted as an integral rotational recoil energy shift. 
It adds up to the  center-of-mass recoil energy 
\begin{equation}
 E_R \rightarrow E_R + \frac{q^2}{2 M } =  \frac{p^2_\beta}{2 M_t } = 
\frac{\epsilon_\beta}{M_t} \left( 1 + \frac{\epsilon_\beta}{2 m_e c^2} \right) .
\label{eq:comp_rec}
\end{equation}
The additional  operator term in (\ref{eq:9}) is estimated at 
$\left| \frac{({\bf q},{\hat{{\bf p}}}_{{\bf R}})}{M} \right| \sim 10^{-2} $. We will use 
it like an expansion parameter. 

To start with $\theta$-function in (\ref{eq:7}), we use following relation:
\[ f(\hat{A}+\hat{\varepsilon}) = f(\hat{A}) + f'(\hat{A})\hat{\varepsilon} + \frac{1}{2}
f''(\hat{A})[\hat{\varepsilon},\hat{A}] + 
O(\hat{\varepsilon}^2, [\hat{A},[\hat{\varepsilon},\hat{A}]],\ldots)
\]    
to get 
\begin{equation}
\theta (\hat{\epsilon} - \frac{({\bf q},{\hat{{\bf p}}}_{{\bf R}})}{M})
 =  \theta (\hat{\epsilon} )  - \delta (\hat{\epsilon})
\frac{({\bf q},{\hat{{\bf p}}}_{{\bf R}})}{M} + \ldots ,
\label{eq:10}
\end{equation}
The general functions must be understood as usually in terms of convolution with 
smooth-class  functions. In our case it is naturally provided by the final energy 
resolution. The experimentally observed spectra are defined by
\[ N_{exp}(\epsilon_\beta) = \int d \epsilon R(\epsilon_\beta - \epsilon) N(\epsilon) ,
\]  
where $N(\epsilon)$ is the one from (\ref{eq:2}). Taking for example generic case 
of Gaussian 
\[ R(\epsilon) = \frac{1}{\sqrt{2\pi\sigma}}\displaystyle{e^{-\epsilon^2/2\sigma^2}}, \]
we conclude that $\delta$-like  terms in (\ref{eq:10}) will produce local corrections the type 
shown in Fig.\ref{fig2} 
Moreover, due to additional $\int d \hat{{\bf q}}$ 
integration leading order correction  will be quadratic in  
$\left| \frac{({\bf q},{\hat{{\bf p}}}_{{\bf R}})}{M} \right|^2$. Their total spectral 
power should be sufficiently small, therefor we keep  in (\ref{eq:10}) leading term 
only. 

It is easy to follow that expansion of "3/2"-function over 
$\frac{({\bf q},{\hat{{\bf p}}}_{{\bf R}})}{M}$ parameter with consequent  
$\int d \hat{{\bf q}}$ average will give:
\begin{equation}
\left( \hat{\epsilon}^2 -  m^2_\nu c^4\right)^{3/2} - 
\left(\frac{q}{M}\right)^2 \Delta_{{\bf R}} \hat{\epsilon} + \hat{C} +
O(\frac{(m_\nu c^2)^4}{\hat{\epsilon}^2} 
\left| \frac{({\bf q},{\hat{{\bf p}}}_{{\bf R}})}{M} \right|^2).
\label{eq:12}
\end{equation} 
The particular form of $\hat{C}$ depends on the operator order in (\ref{eq:12}) and is given by 
\[ \hat{C} = -\frac{1}{3} \left(\frac{q}{M}\right)^2 
\left( \left[\left[H,\frac{d}{dR}\right],\frac{d}{dR}\right] - \frac{d}{dR} \left[H,\frac{d}{dR}\right]\right) .
\]   
The spectral power of this term is relative small, e.g. for $T ^3He^+$ electronic ground 
state we have $|\hat{C}| \le 0.1$ $eV^3$.   
   
Putting all together  we get a new representation for the $\beta$-spectrum:
\begin{equation}
\begin{array}{l} 
N(\epsilon_\beta)  =  \frac{A}{3} F(p_\beta,Z) E_\beta p_\beta 
<T_2| [\left( \hat{\epsilon}^2 -  m^2_\nu c^4\right)^{3/2} - 
 \left(\frac{q}{M}\right)^2 \Delta_{{\bf R}} \hat{\epsilon} + 
 \hfill \hat{C} ]  \theta (\hat{\epsilon} ) |T_2> .  
\end{array} 
\label{eq:13}
\end{equation}  
The differential spectrum is written down analogously:
\begin{equation}
\begin{array}{l} 
\frac{d N(\epsilon_\beta)}{d \epsilon_\beta} =  {A} F(p_\beta,Z) E_\beta p_\beta 
<T_2| [\hat{\epsilon} \left( \hat{\epsilon}^2 -  m^2_\nu c^4\right)^{1/2} - 
 \frac{1}{3} \left(\frac{q}{M}\right)^2 \Delta_{{\bf R}}] \theta (\hat{\epsilon} ) |T_2> .  
\end{array} 
\label{eq:13d}
\end{equation}  
The new final states expansion formula appears after plug-in $\sum_n |n><n|=1$ into 
(\ref{eq:13}), (\ref{eq:13d}). For the integral spectrum we have:
\begin{equation}
\begin{array}{l} 
N(\epsilon_\beta)  =  \frac{A}{3} F(p_\beta,Z) E_\beta p_\beta 
\sum_n \tilde{P}_n [\left( \epsilon_n^2 -  m^2_\nu c^4\right)^{3/2}  - 
\left(\frac{q}{M}\right)^2\epsilon_n <T_2|\Delta_{R}| n><n|T_2>/\tilde{P}_n + \\
\hfill <T_2|\hat{C}| n><n|T_2>/\tilde{P}_n]  \theta (\epsilon_n) \quad , \quad  
\tilde{P}_n = |<T_2|n>|^2 .
\end{array} 
\label{eq:14}
\end{equation}  

\section{Discussion }

Let us discuss derived formulas. Compare formulas (\ref{eq:5c}) and (\ref{eq:13}) 
we get the following operator expressions for the cumulative momenta:
\begin{equation}
\begin{array}{l}   
P_\epsilon  = <T_2|\theta (\hat{\epsilon} )|T_2> \quad , \quad 
<E_n>_\epsilon = <T_2|H \theta (\hat{\epsilon} )|T_2>/P_\epsilon + \frac{q^2}{2 M } \\
<E^2_n>_\epsilon = [<T_2|(H+ \frac{q^2}{2 M })^2\theta (\hat{\epsilon} )|T_2> 
- \left(\frac{q}{M}\right)^2 <T_2|\Delta_{R}| T_2> ]/P_\epsilon .
\end{array} 
\label{eq:15}
\end{equation}  

As it follows from (\ref{eq:14}), the {\it pseudo}-spectrum of final excitations
contains vibrational modes only. The term $\frac{q^2}{2 M }$ represents uniform 
rotational recoil energy shift. Consider as an example $T ^3He^+$ ground electronic state.
The pseudo-spectrum of final excitations is emptied by the overlap with the first 
few vibrational states
\[
\begin{array}{c}
|<T_2|g,v=0>|^2 = 52.2 \% , \quad |<T_2|g,v=1>|^2 = 4.62 \% \\ 
|<T_2|g,v=2>|^2 = 0.45 \%, \quad |<T_2|g,v=3>|^2=0.08 \% \ldots 
 \end{array}
\]
The total sum, as before, is equal to the integral probability 57.4\%. 
Using calculated population probabilities, we can find electronic ground state 
first energy moment          
\[ <E_n>_g = \frac{q^2}{2 M }  + 1/P_g \sum_n \tilde{P}_n E_n  = 1.72 + 0.03 =1.75 \ eV . 
\]
To get exactly same result one has to sum some  hundreds lines in the standard approach.
Note that vibrational recoil shift is generally $\sim 0.1 \ {\rm eV}$. The sign of the shift 
depends  on the particular state. As for the first excited electronic state it is negative.
In the relation to the highly excited electronic states our approach also give clear recipe 
on how to incorporate correctly nuclear motion. It will be enough to consider that 
in the final state nuclei move   under  effective Coulomb repulsion $Z(\approx 2)/R $.

We would like to discuss now one very interesting consequence of our formulas. We have noted 
already in the course our derivation that composite center-of-mass, rotational  recoil shift 
is 
\[ E_R = \frac{\epsilon_\beta}{M_t} \left( 1 + \frac{\epsilon_\beta}{2 m_e c^2} \right)
\approx E^{0}_R - \frac{W_0-\epsilon_\beta}{M_t} .
\]
That corresponds to the effective change of the end-point energy $W_0$ as it is seen
at $ W_0-\epsilon_\beta$ below true end-point energy $W_0$:
\[ W^{(eff)}_0 = W_0 -\delta W_0 \quad , \quad \delta W_0 = \frac{W_0-\epsilon_\beta}{M_t} .
\]

One can guess that $W_0$ fit value will be dominated by the lower energy end, where 
spectral yield is huge. At $\sim 200 \ {\rm eV}$ below end-point energy we have
$ \delta W_0 \sim 0.04 \ {\rm eV}$. This value is comparable with the statistical
uncertainty of $W_0$. Fixing $W^{(eff)}_0$, we sweep into the 
$W_0$ energy region, where small neutrino mass term is comparable with the total 
spectral yield. To compensate  deficit of spectral yield caused by lower value of 
$W^{(eff)}_0$, we  have to make $m^2_\nu$  negative. The rough estimate give us 
some few ${\rm eV^2}$ of negative shift. The fit experiments 
have to show real size of the effect.  
   
The integral recoil energy shift is clearly associated with the $p_\beta$-dependence of 
the final states probability distribution $P_n( p_\beta)$. The precise nature of that 
dependence wasn't available, apart the fact that $P_n( p_\beta)$ change wildly at small
variations of $q(p_\beta)$.

Last notes concern of the other possible experimental situations. The case of $TH$ 
decaying molecule passes through our consideration with obvious changes. To note 
some: the rotational recoil shift $\frac{p^2_\beta}{2 M_t (1+ M_t/M_p)}$ is about 
half of the $T_2$ case; the composite recoil shift (\ref{eq:comp_rec}) stays the 
same. The case of initial nonzero $J$ rotational state leads in final channel to the 
selection of $J$ rotational states, which causes  uniform spectral 
shift $\sim \frac{(J+1/2)^2}{2M R^2_e}$. Finally, there are shouldn't principal
difficulties to make proper  modifications caused by the solid state surrounding.

As a conclusion, we can suggest one modification which is  easy to incorporate 
into standard spectrum. It is slight energy dependence of the end-point energy
\[ W_0  \Rightarrow W_0 -    \frac{W_0-\epsilon_\beta}{M_t} .
\]
This modification is almost irrelevant for the $W_0$ itself but it can sensibly 
change final value of $m^2_\nu$ parameter.

\section{Acknowledgment}

I would like to thank many people who encourage me and help with useful discussions.
The special thanks go to Petr Braun, Gernot Alber and Michael Walter.    
The author acknowledge support of SFB 276 located in Freiburg.



\begin{figure}
\caption{The final states spectrum.
}
\label{fig1}
\end{figure}
\begin{figure}
\caption{The absolut difference between spectral sum and its linear approximation for 
$m_\nu = 1 \ eV$ reference value. The additional line shows 
$({m_\nu} c^2)^4/(W_0-\epsilon_\beta)$ trend line. }
\label{fig2}
\end{figure}

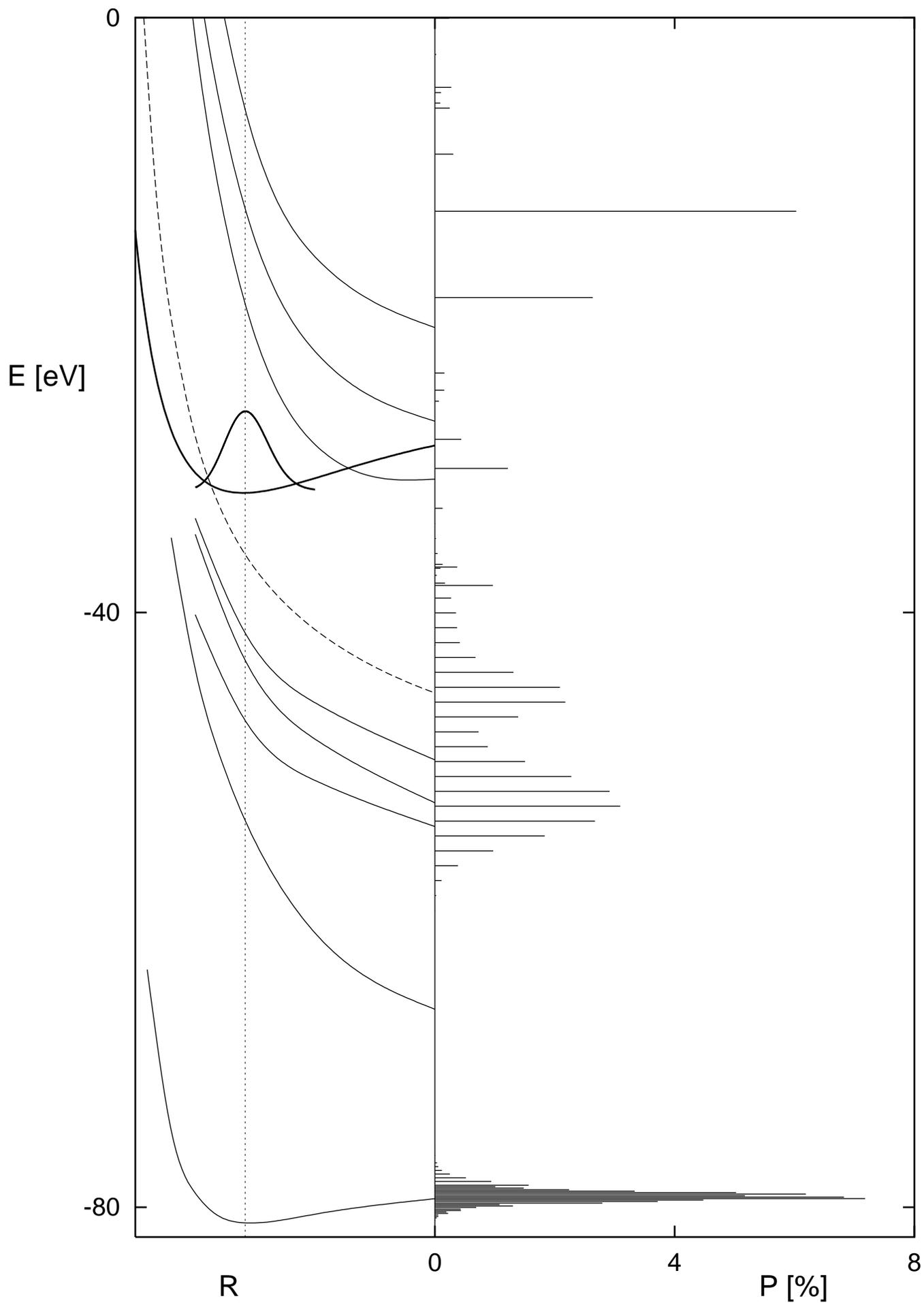
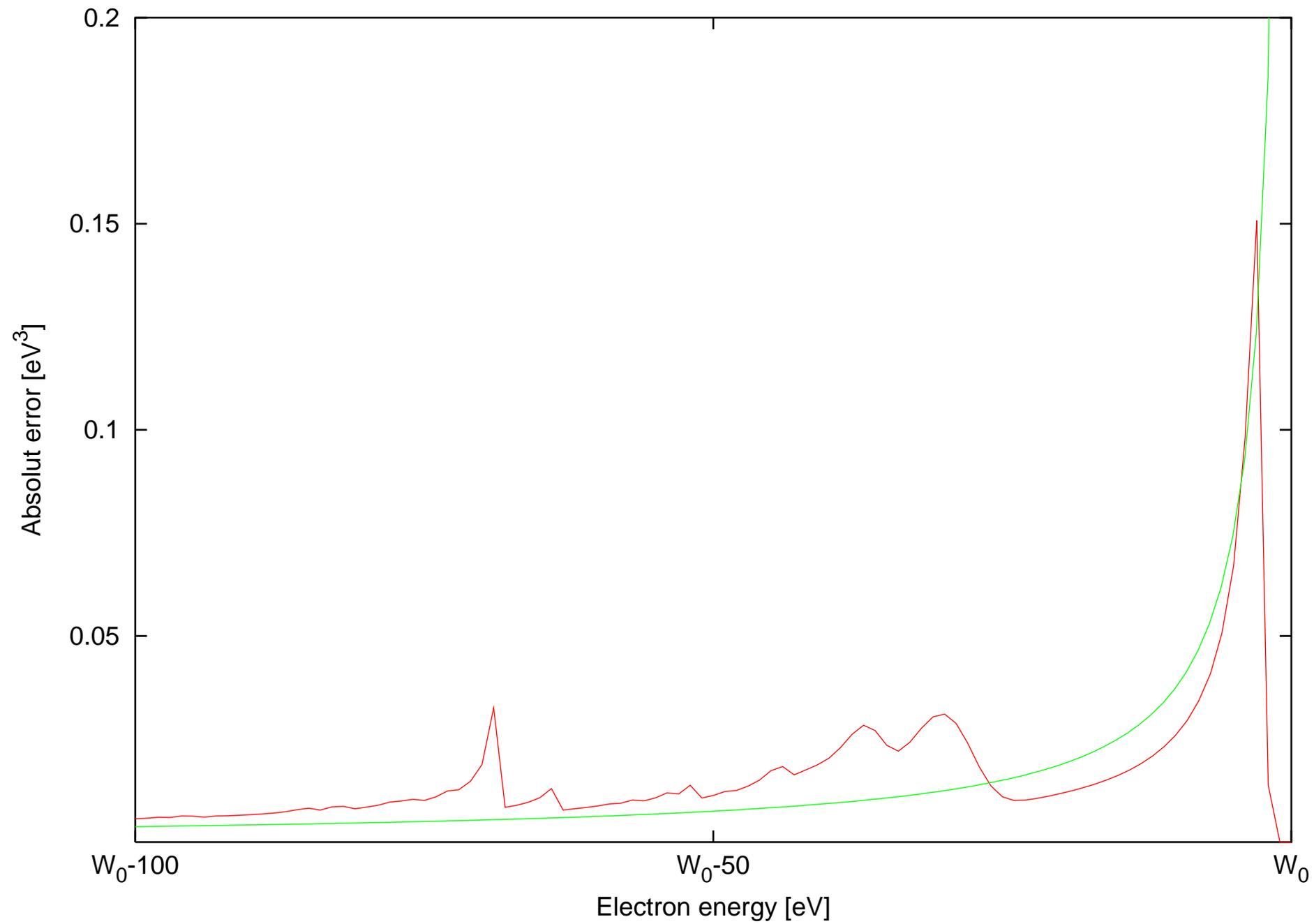
\end{document}